\documentstyle[11pt,cs10,epsf] {article}

\begin {document}

\title {ROTATIONAL MIXING AND THE PRIMORDIAL LITHIUM ABUNDANCE}

\author {M.H. Pinsonneault, V.K. Narayanan, G. Steigman\altaffilmark{1}, 
and T.P. Walker\altaffilmark{1}}

\affil{The Ohio State University, Department of Astronomy}

\altaffiltext{1}{Also Department of Physics}

\begin{abstract}

There has been recent progress in the study of the angular momentum 
evolution of low mass stars (Krishnamurthi et al 1997a).  Theoretical 
models can now be constructed which reproduce the angular momentum 
evolution of low mass open cluster stars and the distribution of 
initial conditions can be inferred from young clusters.  In this 
poster we report on the application of these models to the problem 
of rotational mixing in halo stars.

The distribution of initial conditions inferred from young clusters 
produces a well-defined halo lithium ``plateau" with modest scatter 
and a small population of outliers.  Different choices for the solar 
calibration produce a range of absolute depletion factors.  We show 
that both the dispersion and the ratio of $^6$Li depletion to $^7$Li 
depletion increase as the absolute $^7$Li depletion increases.  The 
measured $^6$Li in HD 84937 and the dispersion in the plateau set 
independent upper bounds on the $^7$Li depletion.  Consistency 
with open clusters and the Sun, along with claims of an intrinsic 
dispersion in the plateau, set a lower bound.  We derive a range of 
0.2-0.4 dex $^7$Li depletion in halo field stars.  Implications for 
cosmology are discussed.

\end{abstract}


\section {Introduction : Rotational Mixing and Lithium}

\begin{itemize}
\item $^7$Li is important as a test of Big Bang nucleosynthesis 
(Boesgaard \& Steigman 1985).  Current estimates of the primordial 
$^4$He favor low baryon densities, while conflicting estimates of 
the primordial deuterium abundance are consistent either with low 
or high baryon densities.  $^7$Li may enable us to distinguish 
between these options.

\item Lithium is destroyed at relatively low temperatures in 
stellar interiors (of order $2.5 \times 10^6$ K).  Surface 
lithium abundances can therefore be affected by nuclear burning 
(if the surface convection zone is sufficiently deep), mass loss, 
microscopic diffusion, and mixing.

\item The lithium depletion pattern seen in open clusters is not 
consistent with the predictions of standard stellar models (see 
Pinsonneault 1997 for a review.)

\item Although the agreement between standard models and halo stars 
is better, implying a low primordial $^7$Li, the $^7$Li abundances 
of halo stars may have been affected by the same process affecting 
open clusters.

\item Rotational mixing is an attractive explanation for this overall 
pattern.  The time dependence of lithium depletion in models with 
rotational mixing matches the data (Chaboyer et al. 1995a,b).  Models 
with different initial rotation rates experience different degrees 
of mixing, providing a natural explanation for a dispersion at fixed 
$T_{eff}$ (Pinsonneault et al. 1989).

\item In this poster we examine the distribution of lithium depletion 
factors expected from the observed distribution of initial conditions 
seen in young cluster stars.  We show that the slope of the Li-$T_{eff}$ 
relationship is consistent with rotational models and use the halo star 
dispersion and the ratio of $^6$Li to $^7$Li depletion in HD 84937 to 
place limits on the primordial $^7$Li.
\end {itemize}

\section {Rotational Properties of the Models}

Internal angular momentum transport from hydrodynamic mechanisms, 
and the associated mixing, is included in the models (Krishnamurthi 
et al 1997a).

The initial conditions are generated by assuming that accretion disks 
enforce a rotation period of 10 days in the surface convection zone 
of a T Tauri star until they decouple from the central star.   A range 
of disk lifetimes is used to generate a range of rotation rates on the 
main sequence.  The distribution of disk lifetimes needed to reproduce 
the observed rotation rates in young open clusters is strongly peaked, 
with a majority of slow rotators and a small population of fast-spinning 
stars (Krishnamurthi et al. 1997a).

Rotational mixing is calibrated by requiring that for some value of 
the solar accretion disk lifetime a solar model have the solar $^7$Li 
abundance at the age of the Sun.  We chose three different initial 
conditions (0, 0.3, and 1 Myr disk lifetimes) for the Sun.  The no 
disk case corresponds to the assumption that the Sun was a rapid 
rotator in its youth (and thus will experience more lithium depletion 
than typical for stars of its mass and age).  The 1 Myr disk case 
would make the Sun a slow rotator, which would imply that the solar 
lithium depletion was typical of its mass and age.

An angular momentum loss law which saturates at high rotation rates 
is used; the observational data requires a saturation threshold that 
depends on the convective overturn time scale (Patten \& Simon 1996, 
Krishnamurthi et al. 1997b).

\section {Lithium Depletion in Open Cluster Stars}

In previous work, rotational mixing from a wide range of initial 
angular momenta has been considered.  In this poster, we use the 
observed distribution of rotation rates in young open cluster 
stars to infer a distribution of initial conditions.  We then 
examine the distribution of lithium abundances that would be 
seen in open cluster stars of different age as follows :

\begin {itemize}
\item  Models with the [Fe/H] and ages of the clusters were 
constructed.  For the $T_{eff}$ of each cluster star with 
lithium data, a disk lifetime was chosen at random from the 
Pleiades distribution of initial conditions.  We looked at 
two different solar calibrations.  In Figure \ref{fig-1} 
simulated lithium depletion patterns for open cluster stars 
are shown for the models where the Sun is at the birthline.

\begin{figure}
\centerline{
\epsfxsize=\hsize
\epsfbox[18 144 592 738]{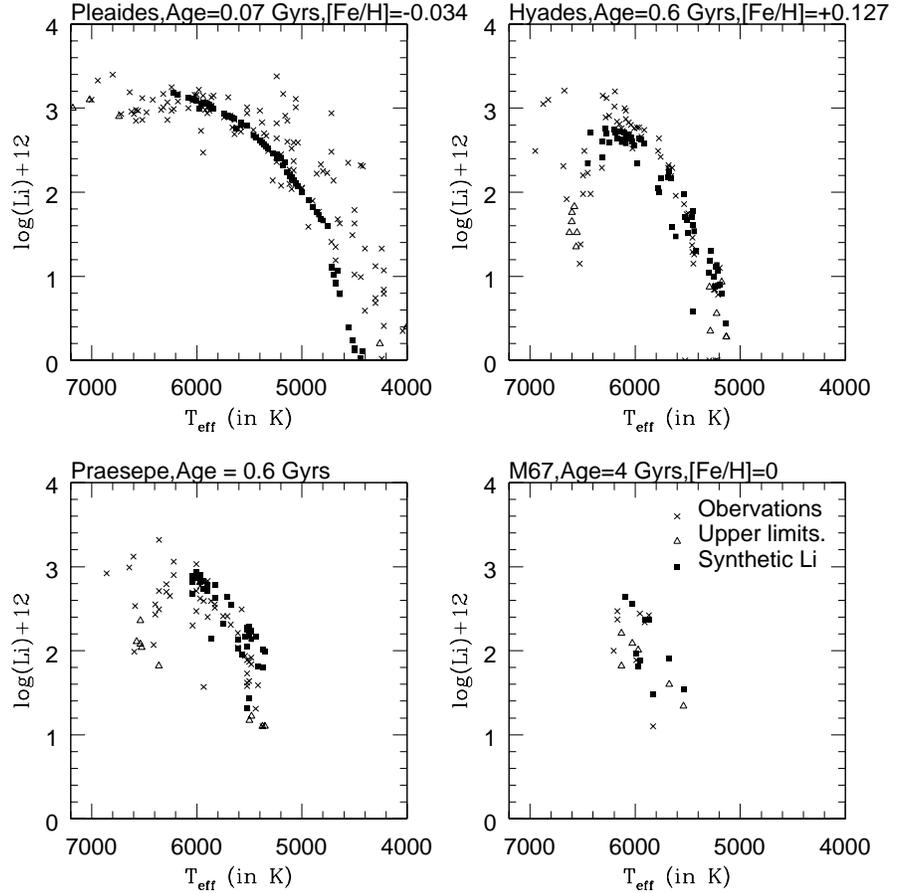}
}
\caption {Simulated distributions of $^7$Li abundances in open 
clusters compared with data (Soderblom et al. 1993, Deliyannis 
et al. 1994, Balachandran 1995).  For each data point, the initial 
condition was randomly drawn from the Pleiades distribution of 
initial conditions.  A lithium depletion appropriate for that 
initial condition was then applied to an assumed initial abundance 
log N(Li) of 3.4 (young systems) or 3.3 (M67 and the Sun) on the 
scale where log N(H) = 12.  The models in this figure used a solar 
calibration in which the Sun was assumed to have a 0 Myr disk 
lifetime; this represents the minimum mixing and dispersion case.} 
\label {fig-1}
\end{figure}

In Figure \ref{fig-2} we show simulated lithium depletion patterns 
for open cluster stars for a model where the Sun is a typical star 
with a 1 Myr disk.

\begin{figure}
\centerline{
\epsfxsize=\hsize
\epsfbox[18 144 592 738]{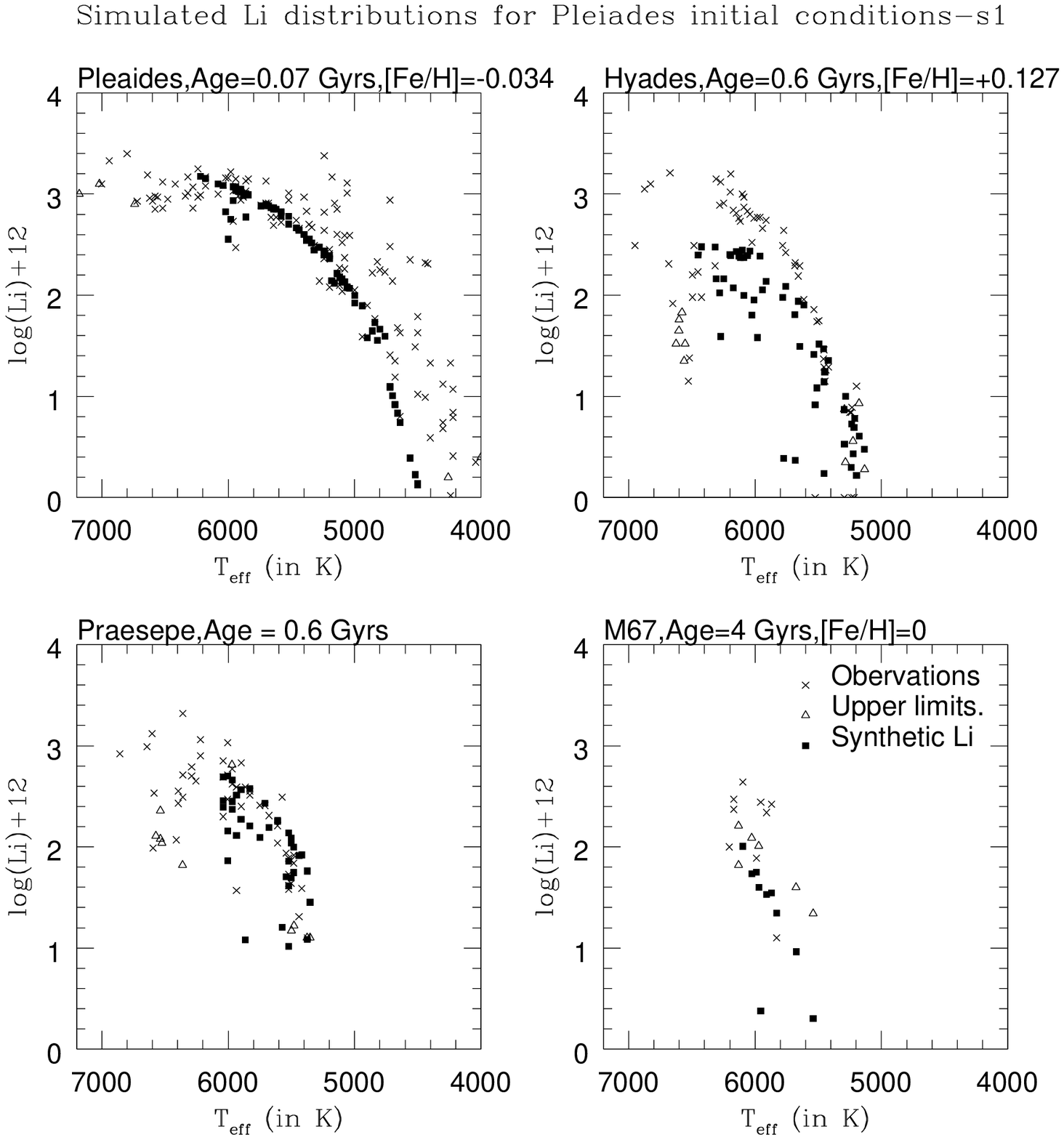}
}
\caption {The same as Figure 1, except that the models in this 
figure used a solar calibration in which the Sun has a 1 Myr disk; 
this represents a large mixing and dispersion case.} \label {fig-2}
\end{figure}

\item  There is minimal rotational mixing in young stars.
\item  A narrow band of depletion factors for the majority 
of open cluster stars is produced by the slow rotators in 
the Pleiades distribution of initial conditions.
\item  Rapid rotators have different properties.  The wide 
range in lithium depletion seen in young cool stars is not 
produced by early rotational mixing even in rapid rotators.  
Rapid rotators also develop excess lithium depletion on the 
main sequence relative to slow rotators.
\item  This relative pattern implies that if the fraction of 
rapid rotators varies from system to system the magnitude of 
the dispersion could also be different in different systems.  
For example, the large dispersion in globular cluster main 
sequence stars relative to field halo stars (Boesgaard et al. 
1997 [B-100] and Thorburn et al. 1997 [B-112]) can be naturally 
explained in the rotational mixing framework by a different 
distribution of initial conditions.

\end {itemize}

\section {Dependence of Lithium Depletion on the Solar Calibration; 
Halo Lithium Abundances}

\begin{figure}
\centerline{
\epsfxsize=\hsize
\epsfbox[18 144 592 738]{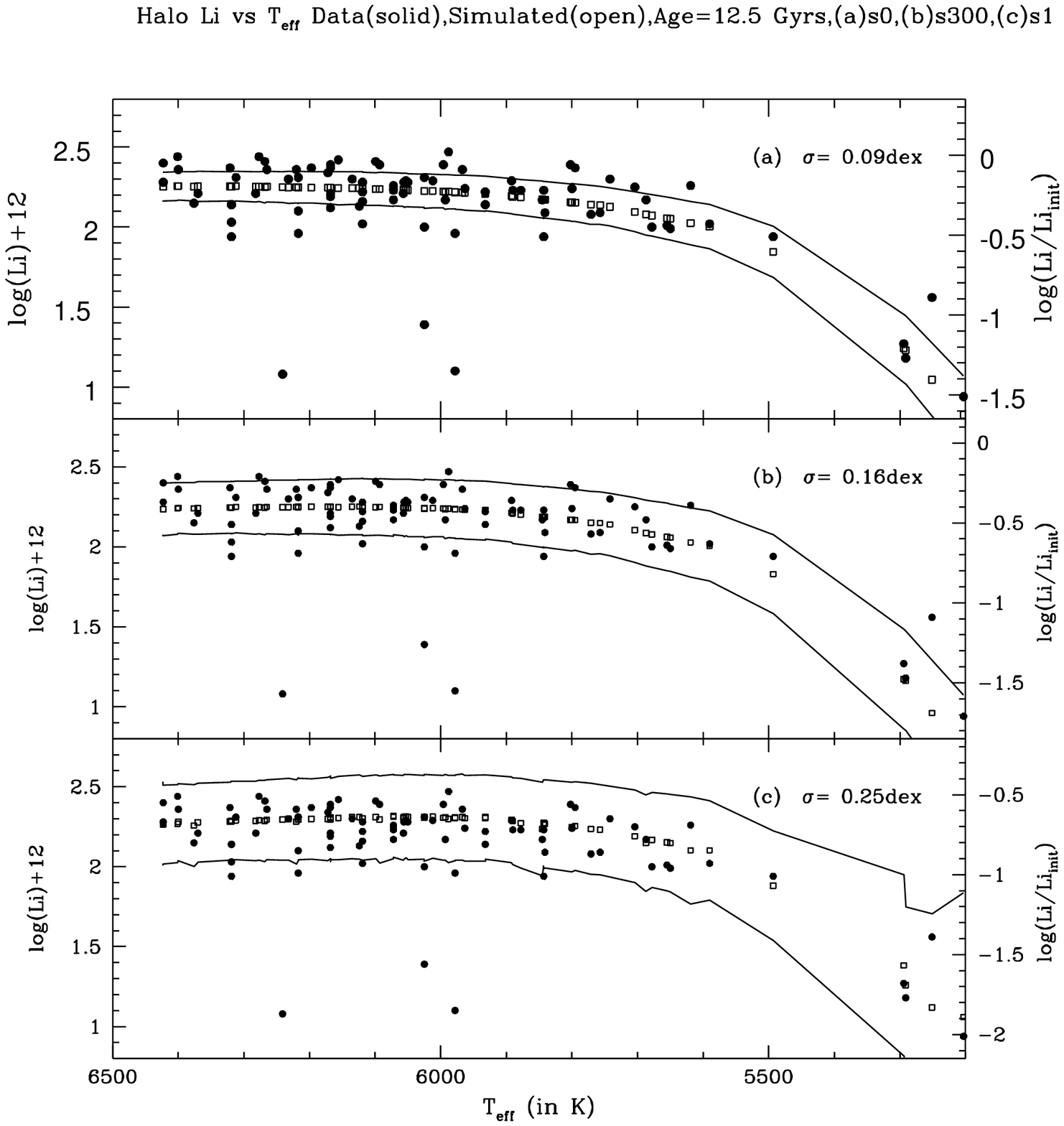}
}
\caption {Lithium abundances for halo stars (data from Thorburn 1995) 
are compared with simulations for three different solar calibrations.  
The open symbols are the mean depletion values for each data point in 
the simulation and the solid lines are displaced +/- 1 $\sigma$ from 
the mean trend in the models.  In panel a, the calibration with the 0 
Myr disk coupling time scale for the Sun is used; a primordial $^7$Li 
abundance log N(Li)$_p$ = 2.5 is inferred.  In panel b, the calibration 
with the 0.3 Myr disk coupling time scale for the Sun is used; log 
N(Li)$_p$ of 2.7 is inferred.  In panel c, the calibration with the 1 
Myr disk coupling time scale for the Sun is used; log N(Li)$_p$ of 3.1 
is inferred.} 
\label {fig-3}

\end {figure}

Simulated distributions of halo lithium abundances for the three 
solar calibrations are shown in Figure \ref{fig-3}.  As the degree 
of mixing increases, both the mean depletion and dispersion increase.

A nearly flat plateau, with little scatter, is expected for the 
distribution of initial conditions seen in young clusters.  The 
curvature in the Li-$T_{eff}$ plane present in earlier models of 
rotational mixing is absent because of the mass dependence of the 
angular momentum loss law.

The detection of $^6$Li has been claimed in one hot halo star 
(HD 84937).  Slow mixing allows simultaneous depletion of $^6$Li 
and $^7$Li while preserving detectable amounts of both.  The ratio 
of $^6$Li to $^7$Li depletion ratio increases with increased $^7$Li 
depletion.  The dispersion and $^6$Li/$^7$Li depletion ratio can 
therefore both be used as diagnostics of the degree of $^7$Li 
depletion in halo stars.

\section {Bounds on the Primordial $^7$Li Abundance}

The dispersion in the data of Thorburn (1994) is 0.13 dex around 
the mean Li-$T_{eff}$ trend and 0.16 dex when defined in the same 
way as in the models.  Observational error and a range in metallicity 
and age will compose some of this dispersion, implying that the 
dispersion which can be attributed to rotational mixing will be 
smaller than 0.16 dex.  The dispersion in the halo plateau therefore 
sets an upper bound on the $^7$Li depletion in halo plateau stars of 
0.4 dex.

There is a claimed detection of $^6$Li in one halo star, HD 84937.  
A bound on the initial $^7$Li in this star can be made from the 
$^6$Li/$^7$Li ratio if a maximum initial $^6$Li abundance can be 
established.  $^6$Li is produced by spallation, either from CNO 
nuclei or from $\alpha - \alpha$ fusion synthesis.  Stringent 
limits of 0.6 dex $^6$Li depletion have been claimed based on 
CNO spallation (Lemoine et al. 1997).  However, the observed 
level of $^6$Li is higher than the predicted level even before 
possible depletion is taken into account.  If $\alpha - \alpha$ 
fusion synthesis is included we can set a firm limit on the 
$^6$Li depletion of 1.3 dex, corresponding to an initial $^6$Li 
for HD 84937 equal to the solar meteoritic value.  A limit of 
0.55 dex $^7$Li depletion can be placed based on the $^6$Li 
constraint.  This limit is less stringent than the limit based 
on the dispersion.

Consistency with the open cluster data, the inferred intrinsic 
dispersion in the plateau, and the existence of highly overdepleted 
halo stars all act to bound the $^7$Li depletion from below.  The 
lower bound on the $^7$Li depletion is 0.2 dex.

\section {Cosmological Implications}

An initial abundance of $2.25 \pm 0.1$ and a depletion of $0.2 - 0.4$ 
dex produces a predicted range 2.35 $<$ log N(Li)$_p$ $<$ 2.75, or

$2.2 < 10^{10}(Li/H) < 5.6$.

There are two ranges of the baryon to photon ratio $\eta$ consistent 
with the above limits on log N(Li)$_p$ : a ``low $\eta$" branch with 
$0.8 < \eta_{10} < 1.7$ and a ``high $\eta$" branch with
$3.7 < \eta_{10} < 9.0$ ($\eta_{10}$ = $\eta$ in units of $10^{-10}$)

The inferred log N(Li)$_p$ is in good agreement with the $\eta$ implied 
by low deuterium abundances in QSOs (Tytler et al. 1996) for the high 
$\eta$ branch.  This implies

$0.014 < \Omega_bh^2 < 0.033$.

The inferred log N(Li)$_p$ is in good agreement with the $\eta$ inferred 
from $^4$He (Olive et al. 1997) for the low $\eta$ branch and consistent 
with the high deuterium abundances claimed for some QSOs (Rugers \& Hogan 
1996).  The baryon density corresponding to this branch is low,
 
$0.003 < \Omega_bh^2 < 0.006$.

\begin {references}
\reference Balachandran, S. 1995, \apj, 446, 203
\reference Boesgaard, A.M. \& Steigman, G. 1985, \araa, 23, 319
\reference Boesgaard, A.M., Deliyannis, C.P., Stephens, A., \&
King, J.R. 1997, \cswten, in press [B-100]
\reference Chaboyer, B., Demarque, P., \& Pinsonneault, M.H. 1995a,\apj, 441, 
865
\reference Chaboyer, B., Demarque, P., \& Pinsonneault, M.H. 1995b,\apj, 441, 
876
\reference Deliyannis, C.P., King, J.R., Boesgaard, A.M., \& Ryan, S.G. 1994,
\apjl, 434, L71 
\reference Krishnamurthi, A., Pinsonneault, M.H., Barnes, S., \& Sofia, S. 
1997a, \apj, 480, 303
\reference Krishnamurthi et al. 1997b, \apj, in press
\reference Lemoine, M., Schramm, D.N., Truran, J.W., \& Copi, C.J. 1997,
\apj, 478, 554
\reference Olive, K.A., Steigman, G., \& Skillman, E.D. 1997, \apj, 483, 788
\reference Patten, B.M. \& Simon, T. 1996, \apjs, 106, 489
\reference Pinsonneault, M.H., Kawaler, S., Demarque, P., \& Sofia, S. 1989,
\apj, 338, 424
\reference Pinsonneault, M.H. 1997, \araa, 35, 557
\reference Rugers, M. \& Hogan, C.J. 1996, \aj, 111, 2135
\reference Soderblom et al. 1993, \aj, 106, 1059
\reference Thorburn, J.A. 1994, \apj, 421, 318
\reference Thorburn, J.A., Deliyannis, C.P., Rubenstein, E.P., Rich, R.M.,
\& Orosz, J.A. 1997, \cswten, in press [B-112]
\reference Tytler, D., Fan, X.-M., \& Burles, S. 1996, Nature, 381, 207
\end {references}

\index{Sun}
\index{Pleiades}
\index{Hyades}
\index{HD 84937}
\index{Praesepe}
\index{M67}

\end {document}